\documentclass[prl,twocolumn]{revtex4}
\usepackage{amsfonts}
\usepackage{amsmath}
\usepackage{amssymb}
\usepackage{graphicx}

\begin{document}

\title{\bf Comment on "Fermi-Bose Mixtures near Broad Interspecies Feshbach Resonances"}
\maketitle


In a recent Letter, Song {\it et al.} [1] introduced a new variational approach to treat
strong attractive boson-fermion (BF) correlations in BF atomic mixtures. The proposed theory predicts a first order phase
transition to a condensate of composite BF pairs with center of mass momentum $\mathbf{Q}=0$ as opposed to a composite fermionic molecular Fermi gas.
We will show in this comment that their approach is incorrect and moreover, by resorting to
an exactly solvable model we will demonstrate that there cannot be more than
one correlated $\mathbf{Q}=0$ BF pair in complete
contradiction with the conclusions of [1].

Let us start with the mean-field BF Hamiltonian restricted to $\mathbf{Q}=0$ BF pairing as
considered in [1]

\begin{equation}
H=\sum_{\mathbf{k}}\epsilon _{\mathbf{k}}^{b}b_{\mathbf{k}}^{\dagger }b_{%
\mathbf{k}}+\sum_{\mathbf{k}}\epsilon _{\mathbf{k}}^{f}f_{\mathbf{k}%
}^{\dagger }f_{\mathbf{k}}+g\sum_{\mathbf{kk}^{\prime }}f_{\mathbf{k}}^{\dagger }b_{-\mathbf{%
k}}^{\dagger }b_{-\mathbf{%
k'}}f_{\mathbf{k'}},  \label{Ha}
\end{equation}%
 with the same notation as in Ref. \cite{1}.
 In the evaluation of the expectation value of the interaction part in the variational  state (5) of
[1] special care has to be taken with the fermionic anticommutation relations. It is then straightforward to obtain $\left\langle \text{g.s.}\right\vert f_{\mathbf{k}}^{\dagger }b_{-\mathbf{%
k}}^{\dagger }b_{-\mathbf{%
k'}}f_{\mathbf{k'}}\left\vert \text{g.s.}%
\right\rangle _{\left( k<k^{\prime }\right) }=(1-\eta_{\mathbf{k}}) (1-\eta_{\mathbf{k'}}) u_{\mathbf{k}}v_{\mathbf{k}}u_{%
\mathbf{k}^{\prime }}v_{\mathbf{k}^{\prime }}{\prod\limits_{\mathbf{k}^{^{\prime
\prime }}\left( \mathbf{k}<\mathbf{k}^{\prime \prime }<\mathbf{k}^{\prime }\right) }}\left( u_{\mathbf{k%
}^{\prime \prime }}^{2}-v_{\mathbf{k}^{\prime \prime }}^{2}-\eta_{\mathbf{k}^{\prime \prime }}^{2}\right) $. We note
here that the string factor, which is directly related to the Pauli principle, should have been missed in Ref. [1] in order to derive their BCS-like equations.
In what follows we will show that the defective mean field approach of [1] led the authors to wrong conclusions.

Instead of proceeding with the correct variational approach, we note here that the Hamiltonian (\ref{Ha}) is
 exactly solvable with eigenstates similar to those of the Richardson exact solution of the BCS model \cite{2}

\begin{equation}
\left\vert \Psi \right\rangle =\prod_{\alpha =1}^{M}\Gamma _{\alpha
}^{\dagger }\left\vert \nu^{b}\nu^{f}\right\rangle , \Gamma _{\alpha
}^{\dagger }=\sum_{\mathbf{k}}\frac{1}{\varepsilon _{\mathbf{k}}-e_{\alpha }}f_{\mathbf{k}}^{\dagger }b_{-\mathbf{%
k}}^{\dagger },\varepsilon _{\mathbf{k}}=\epsilon _{\mathbf{k}}^{b}+\epsilon
_{\mathbf{k}}^{f}  \label{WF}
\end{equation}%
where $M$ is the number of BF pairs and $e_{\alpha }$ are the
spectral parameters (pair energies) and $\nu^{b},\nu^{f}$ are the seniorities ( {\it{i.e.}} the number
of unpaired bosons or fermions respectively in each single particle state, $\left\vert \nu \right\rangle \equiv \left\vert \nu_{\mathbf{k}_1}, \nu_{\mathbf{k}_2},\cdots \right\rangle$). Inserting this ansatz in the eigenvalue equation $H\left\vert \Psi \right\rangle =E\left\vert \Psi \right\rangle$ we derive the equation for the pair energies

\begin{equation}
\sum_{\mathbf{k}} \frac{\left( 1+\nu_{\mathbf{k}}^{b}-\nu _{\mathbf{k}}^{f}\right)} {\varepsilon_{\mathbf{k}}-e_{\alpha }}=-1/g,
\label{Eq}
\end{equation}
and the eigenvalues $E=\sum_{\alpha }e_{\alpha }+\sum_{\mathbf{k}}\left(
\epsilon _{\mathbf{k}}^{b}\nu _{\mathbf{k}}^{b}+\epsilon _{\mathbf{k}}^{f}\nu _{\mathbf{k}}^{f}\right) $.

Eq. (\ref{Eq}) is precisely the equation for the eigenvalues $e_{\alpha}$ of a single BF pair in the presence of $\nu^{b}$ unpaired bosons and $\nu^{f}$ unpaired fermions.  The fact that there is no interaction among composite fermions is directly related to the cancellation of the second
commutator-anticommutator $\left\{ \left[ H,\Gamma _{\alpha }^{\dagger }\right], \Gamma
_{\beta }^{\dagger }\right\} =0$ in the evaluation of the eigenvalue equation. The exact eigenvectors of (\ref{Ha}) are completely defined by a configuration of seniorities $\nu^b,\nu^f$ and a set of $M=N^b-N^b_\nu = N^f - N^f_\nu$ pair energies solution of (\ref{Eq}), with $N^{b,f}$ the total number of particles and $N^{b,f}_\nu=\sum_{\mathbf{k}} \nu_{\mathbf{k}}^{b,f}$  the number of unpaired particles of each kind. Eq. (\ref{Eq}) is equivalent to the Random Phase Approximation of a BF pair which has a unique collective solution with $e_{0}<\varepsilon _{0}$.
All other pair energies are non collective roots constrained to
the intervals between successive active single particle energies  $\varepsilon _{\mathbf{k}}$, defining
quasi-free BF pairs. Therefore, no condensation of collective BF pairs with $\mathbf{Q}=0$ is possible. In this respect, the analysis in Ref. [31] in [1] on the fact that a zero energy BF  mode at  $\mathbf{Q}=0$ does not signal an instability was appropriate, albeit criticized in [1]. This is because in $T$-matrix approach a fermionic mode, contrary to bosonic ones, does not become unstable.

The lowest energy solution of the
Hamiltonian (\ref{Ha}) for a mixture with equal numbers of bosons and
fermions $N^{b}=N^{f}=N$ corresponds to $\nu _{0}^{b}=N-1$, $\nu _{\mathbf{k}}^{f}=1$
for $0<k\leq k_{F}$ and $\ M=1$, $i.e.$ the ground state is a Bose
condensate of  $N-1$ bosons in $\mathbf{k}=0$, times a Fermi sea of  $N-1$ fermions with a
hole in $\mathbf{k}=0$, plus a single bound BF pair with binding energy $e_{0}<0$. This conclusion is independent of the cutoff required to renormalize the interaction (\ref{Ha}).

The pairing terms with finite center of mass momentum $\mathbf{Q}$, neglected in [1], will induce correlations between the BF pairs,
leading to a mixture of condensed bosons, free fermions, and a fraction of correlated BF pairs with different center of mass momenta.
Eventually,  in the strong coupling regime there will be a transition to a Fermi gas of heteronuclear molecules in contradiction with the conclusions of Ref. [1] based on an incorrect approach. Whether this is a true phase transition or a smooth crossover is still an open question.

This work is supported by the Spanish Ministry of  Science and Innovation, project number
FIS2009-07277

\medskip

\noindent J. Dukelsky$^1$, C. Esebbag$^2$, P. Schuck$^3$, and T. Suzuki$^4$.\\
\indent $^1$Instituto de Estructura de la Materia, CSIC,
\indent Serrano 123, 28006 Madrid, Spain \\
\indent $^2$Departamento de Matematicas, Universidad de
\indent Alcala, 28871 Alcala de Henares, Spain\\
\indent $^3$Institut de Physique Nucleaire, CNRS and Universit\'e \\
\indent de Paris-Sud, F-91406 Orsay Cedex; and LPMMC,\\
\indent  CNRS and UJF, 38042 Grenoble Cedex 9, France\\
\indent $^4$ Department of Physics, Tokyo Metropolitan
\indent University, Hachioji, Tokyo 192-0397, Japan

\medskip


\end{document}